\documentclass{cimento}
\usepackage{epsf}
\title{A Study of Sequence Distribution \ETC of a Painted Globule
	as a Model for Proteins with Good Folding
	Properties\thanks{{\bf Paper presented at the International
	Conference on Morphology and Kinetics of Phase Separating
	Complex Fluids, Messina, Italy, June 24-28, 1997.}
        {\bf Published in: Il Nuovo Cimento. Vol. 20D, No 12bis, pp. 2383-2391 (1998). } Typos of the original publication corrected by E.G.T.}}
\author{M-T.~Kechadi\from{ins:1},
	R.G.~Reilly\from{ins:1}, 
	K.A.~Dawson\from{ins:2},\\
	Yu.A.~Kuznetsov\from{ins:2}, \atque
	E.G.~Timoshenko\from{ins:2}\thanks{E-mail: Edward.Timoshenko@ucd.ie. Web page: http://darkstar.ucd.ie.}}
\instlist{\inst{ins:1}Department of Computer Science, University College Dublin, Belfield, Dublin 4, Ireland.
\inst{ins:2}Department of Chemistry, University College Dublin, Belfield, Dublin 4, Ireland.
}
\PACSes{
\PACSit{07.05.T}{Computer modeling and simulation}
\PACSit{36.20}{Proteins}
\PACSit{02.50.N}{Monte Carlo method}
\PACSit{01.30.C}{Conference proceedings}
}
\begin{document}

\maketitle

\begin{abstract}
In this paper we present a method to study the folding structure of a
simple model consisting of two kinds of monomers, hydrophobic and
hydrophilic. This method has three main steps: an efficient
simulation method to bring an open sequence of homopolymer to a folded
state, the application of a painting method called {\tt regular hull}
to the folded globule and the refolding process of the obtained
copolymer sequence. This study allows us to suggest a theoretical
function of disorder distribution for copolymer sequences that give
rise to a compacted and well micro-phase separated globule.
\end{abstract}

\section{Introduction}

Proteins are made up of elementary building blocks - 20 different
amino acids. Once synthesized, the protein chain folds into a unique
3-dimensional shape, determined solely by the amino acid primary
structure. The equilibrium folding is a free energy minimization
process that depends on interactions among amino acids. Once folded, a
protein is usually a compact globule. The compactness of the globule
is maintained by the hydrophobic effect, so that the hydrophobic units
are mainly located inside the globule and the hydrophilic ones on the
surface. These hydrophilic units screen the hydrophobic units thereby
preventing aggregation in solution. Although this phase separation
feature is well understood, biologists can neither accurately predict
the folded protein shape for a given primary sequence, nor which
sequences will fold and be stable, rather than aggregate.

In this paper we present a method to study the folding properties of a
model of proteins containing only two kinds of amino acids; hydrophobic
and hydrophilic. This model is referred to as the AB-model. The idea
is to try to find the distribution of disorder for copolymer sequences
that give rise to a compact globule expressed as hydrophobic core and
hydrophilic exterior.
 
The study rests on two complementary methods, the painting method and
artificial neural networks (ANN) method. The painting method is
applied to quite short sequences and the ANN method to long sequences.
The first is necessary to train the ANN. In other words the ANN needs
the training process (which can be done by the painting method) to
study longer sequences which cannot be characterized by the painting
method. At first we only consider short sequences to prove the
validity of the painting method and the applicability of windowing
technique of ANN to the problem of polymer collapse. 

The paper is organized as follows: The next section describes the
model to study the goodness of a sequence for the purposes of folding.
The description of the method of simulations used to collapse a
homopolymer is given in section 3. The painting method applied to a
collapsed homopolymer globule is presented is section 4. Section 5
describes a study of the distribution of disorder in sequences of a
generic model of protein expressed in terms of hydrophobic and
hydrophilic units. This is done by calculating the correlation
function of monomers along the chain. The choice of using the ANN
method is also discussed in the same section. We conclude in section
6.

\section{The Model}

The model presented here rests mainly on the principle that when a
protein folds it turns into a globule, so that predominantly 
the hydrophobic units constitute the core of the globule and the
hydrophilic units the surface. We therefore wish to create an ensemble
of condensed chains all of which possess a hydrophobic core and
hydrophilic exterior with fixed sizes. Each chain sequence of this
ensemble is then known to possess at least one acceptable folded
state. To produce the painting structure of hydrophobic-hydrophilic
monomers we proceed in the following manner:

\begin{enumerate}
  \item Consider an open sequence of a homopolymer of a fixed size,

  \item We perform numerical simulations based on the method described in
        the following section to bring the system to its folded state.

  \item The globule shape depends on the position of each monomer in
        the sequence and the interactions between them. This globule
        is not always compact. Testing the compactness of the globule
        becomes necessary as only the spherical globules are considered.

  \item The core of the globule is painted. The volume of the colored
        core is defined by the hydrophobicity ratio along the chain.
        The painting technique is described in section \ref{sec4}.

  \item We can now ask if all the sequences, each of which has, a good
        folded state, can be refolded from an open conformation to
        that state. Thus, the obtained colored sequences are
        considered as an AB-model consisting of two kinds of monomers
        A and B. The Monte Carlo simulation method is again used for
        copolymer sequences to discriminate the good sequences from
        the bad ones. A sequence is considered to be good if it
        refolds efficiently and bad otherwise. The colored sequences
        are also used to train the ANN.
\end{enumerate}

The primary concern in this paper is to identify any implied
hydrophilic-hydrophobic correlations created by having a hydrophobic
core structure. Thus, we study the distribution of the AB-model. A
sequence of monomers of length N can be described by the binary
variables $\lambda_1, \cdots, \lambda_N$. Without loss of generality
we consider $\lambda_i = +1$ for hydrophilic and $-1$ for hydrophobic.
$\Lambda$ is a random variable and its probability distribution
function can be deduced from the averages $M_k$ defined over the set
$\{\lambda_{m_{1}}, \cdots,\lambda_{m_{k}}\}$. The random variables
$\lambda_1, \cdots, \lambda_N$ are mutually independent. The cumulant
$M_2$ is given by

\begin{eqnarray}
M_2(m_1,m_2) = \gamma_{m_{1}m_{2}} = \langle\lambda_{m_{1}}
\lambda_{m_{2}}\rangle - \langle \lambda_{m_{1}}\rangle\langle
\lambda_{m_{2}} \rangle
\end{eqnarray}

\begin{figure}
\vspace{0.02cm}
\centerline{
\epsfxsize 11cm
\epsffile{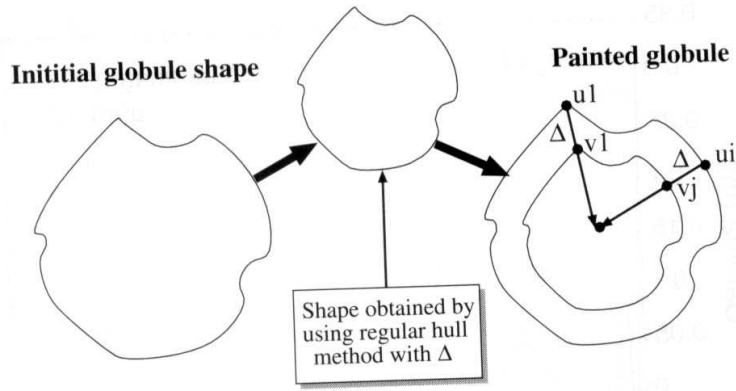}}
\caption{Regular Hull painting method.}
\label{globule}
\end{figure}

\section{Method of the Simulation}

There are two approaches commonly used for computer simulations of
polymer systems. One can proceed by straightforward numerical
integration of, for example, the Langevin equation \cite{4}, or Newton's
equation in the molecular dynamics method. Alternatively, one can
apply the method of Monte Carlo simulation \cite{1,2,3}. The complete
description of the model and simulation methods is given in \cite{5}
based on the package {\tt many\_cop} developed by Yu.A. Kuznetsov.

There are two obvious restrictions on the set of all possible updates
or moves of the system. Namely, we must ensure polymer connectivity,
and excluded volume. In a continuous--space model one requires a
calculation of all forces to ensure that excluded volume is preserved,
and there is an inner ``space'' loop in the Monte Carlo code. This can
be avoided in a model with a finite--size discrete space, since a
look--up table is used to manage this procedure. The dynamics can be
performed by permutations of monomer and solvent beads on the lattice.
We call such a permutation an elementary move.

We consider a model of a copolymer consisting of only two different
monomer types distributed in a certain way along the chain. The total
number of each monomer type is held fixed for every configuration in
the ensemble. The chain structure does not change under time
evolution.

We work on a three-dimensional lattice with unit spacing. We restrict
our model by making the following particular choices of elementary
moves. The maximum distance between the nearest neighbors along the
chain (NNC) is equal to $r_{max} = \sqrt{3}$. Thus, for every bead the
NNC are located in the nearest lattice sites along the vertices of the
lattice, or on second or third lattice neighbors. This condition
provides for connectivity of the chain. Furthermore, excluded volume
is incorporated by ensuring that only NNC are permitted in the nearest
neighbor lattice sites, i.\ e.\ the minimum distance between beads is
$r_{min} = 1$ for NNC beads (NNC cannot overlap), and $r_{min} =
\sqrt{2}$ otherwise.

\def\Interact#1#2{{\cal I}_{#1 #2}}

The  model discussed above is described by the Hamiltonian,
\begin{equation}
   H = \frac{1}{2} \sum_{i \ne j} w(r_{ij})
         \Interact{s_{i}}{s_{j}},
         \label{hamil}
\end{equation}

where $i$, $j$ enumerate lattice sites; $s_{i}$ labels the state of
site $i$, $\Interact{s_{i}}{s_{j}}$ is a 3x3 symmetric matrix and the
matrix indices $s_{i}$ take three different values, solvent $s$ and
monomer types denoted as $a$ and $b$. Here we denote $r_{ij} =
\vert{\bf r}_{i} - {\bf r}_{j}\vert$. For short--range interactions we
take the weight function $w(r) = 0$, for $r > R_{max}$, where
$R_{max}$ is some range of interaction. As in Ref. \cite{5} we choose
$w(1)=1$, $w(\sqrt{2}) = 1$, $w(\sqrt{3}) = 0.7$, $w(2) = 1/2$ and
$w(r) = 0$ for $r > 2$. Thus, the range of interaction includes the
nearest and second--nearest neighbors. We have used the Metropolis
algorithm \cite{1,2,3} for calculation of the transition probability in a
system at temperature $T$.

Copolymers can be described by three independent Flory parameters:
\begin{eqnarray}
 \chi_{aa} & = & \frac{2 \Interact{s}{a} - \Interact{a}{a} -
                 \Interact{s}{s}}{k_{B}T}, \nonumber \\
 \chi_{bb} & = & \frac{2 \Interact{s}{b} - \Interact{b}{b} -
                 \Interact{s}{s}}{k_{B}T}, \label{CPintpar} \\
 \chi_{ab} & = & \frac{  \Interact{s}{a} + \Interact{s}{b} -
                 \Interact{a}{b} - \Interact{s}{s}}{k_{B}T}. \nonumber
\end{eqnarray}
In fact, we shall consider only a special cut of parameter space with
the condition, $\Interact{a}{a} + \Interact{b}{b} = 2\Interact{a}{b}$.
We can therefore reduce the number of parameters to two via the
relation, $\chi_{ab} = (\chi_{aa} + \chi_{bb})/2$. We further restrict
our model by assuming that the $a$--monomers are hydrophilic,
$\chi_{aa} = 0$.

\section{Method of Painting}
\label{sec4}

The painting method was used to identify the two types of monomer -
hydrophobic and hydrophilic. For a given hydrophobicity percentage the
method consists in coloring the interior of the globule with a radius
corresponding to the hydrophobicity ratio $\tau_b$. For this we
consider that the globule is spherical with radius $(R_g \pm
\delta_r)$ with $\delta_r < \epsilon_r$, where $\epsilon_r$ is the
maximum value, and is called {\em parameter of compactness}.

There are different ways to implement the painting procedure,
depending on the constraints to be satisfied. For example, if the
globule is considered a sphere, the easiest way is to mark hydrophobic
all the monomers in a fixed radius $R_b$ which corresponds to the
given amount of hydrophobicity. $R_b$ is called the hydrophobic radius
and the difference $(R_g-R_b)$ is the depth of painting.

Another method consists in bringing the centre of mass of the globule
to the origin of a 3-D Cartesian grid. Calculate the coordinates of
the hydrophobic radius along the 3 axes (x, y, z). This method is
quicker than the first; marking process consists of a simple
coordinate test for each unit compared to those in the radius of
hydrophobicity.

\setlength{\unitlength}{0.62mm}
\begin{figure} 
\vspace{0.02cm}
\begin{picture}(160,160)
\thicklines
\epsfxsize=90mm
\put(0,5){\epsffile{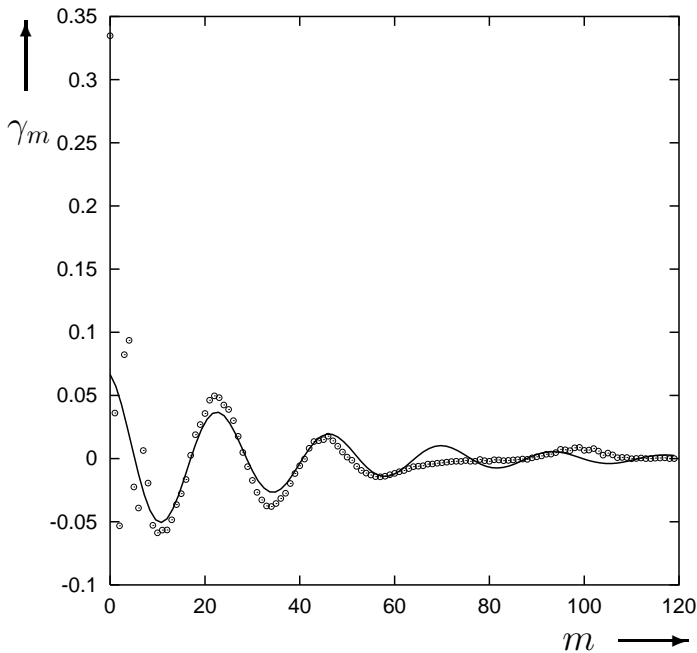}}
\put(118,0){{\Large $m$}}
\put(130,2){{\Large \vector(1,0){15}}}
\put(-1,110){{\Large $\gamma_m$}}
\put(3,120){{\Large \vector(0,1){15}}}
\end{picture}
\caption{Correlation function $\gamma_m = \frac{1}{N-m} \sum_{n}
\gamma_{n,n+m}$ for $\tau_b = 10\%$ and sequence length $N=120$ (circles). 
The theoretical $g(m)$ plot (solid curve) 
corresponds to $A=0.07$, $\xi = 37\pm 5$, $d=23.6 \pm 0.4$ and
$\phi=90$.}
\label{fig2}
\end{figure}

The advantage of these two methods is their simple implementation.
However the constraint imposed on the form of the globule is very
restrictive, especially for short sequences, since: 1) a collapsed
globule is never a perfect sphere and 2) the coordinates of monomers
in the lattice are integers. These approximations cause anomalies in
the final form of the globule.

The method of painting we have chosen involves the {\tt regular hull}.
It consists of defining an internal volume of the same form as that of
the globule. The method proceeds in two phases: 1) determining the
units which constitute the surface of the globule $\Gamma_g =
\{u_0,u_1, \cdots, u_n\}$. 2) For each $u_i\in \Gamma_g$ its distance
is calculated from the centre of mass of the globule $R_{u_{i}}$. The
depth of painting $\Delta$ is fixed, the distance separating a unit
$v_{j}$ of a contour in the internal volume $\Gamma_p
=\{v_0,v_1,\cdots, v_m\}$ from the centre of mass is given by the
equation $\Delta = R_{u_{i}} - R_{v_{j}}$, so the line $[u_i,v_j]$
passes through the centre of mass. The hydrophobic units are therefore
delimited by $\Gamma_p$. Figure \ref{globule} illustrates this
procedure. Note that as we are working on a lattice, the unit
$u_i\in\Gamma_g$ corresponding to $v_{j}$ is chosen in such a way so
as to guarantee the painting depth distance.

\section{Simulations}

We performed Monte Carlo simulations of systems without reptation for
both folding and refolding processes. The simulations were run on
workstations (DEC Alpha 3100 and SGI R10000). The method has two
important steps which are very time consuming - Monte Carlo
simulations and neural network windowing method. For each Monte Carlo
simulation run more than $N^2$ sweeps were carried out, which requires
approximately $N^2$ seconds on the underlined machines. For $N=120$, 4
hours CPU time required to bring the initial sequence to folding
state. Due to the enormous amount of time needed to simulate sequences
of different size, we only ran simulations for $N=120, 250,$ and $400$.

\setlength{\unitlength}{0.62mm}
\begin{figure} 
\vspace{0.02cm}
\begin{picture}(160,160)
\thicklines
\epsfxsize=90mm
\put(0,5){\epsffile{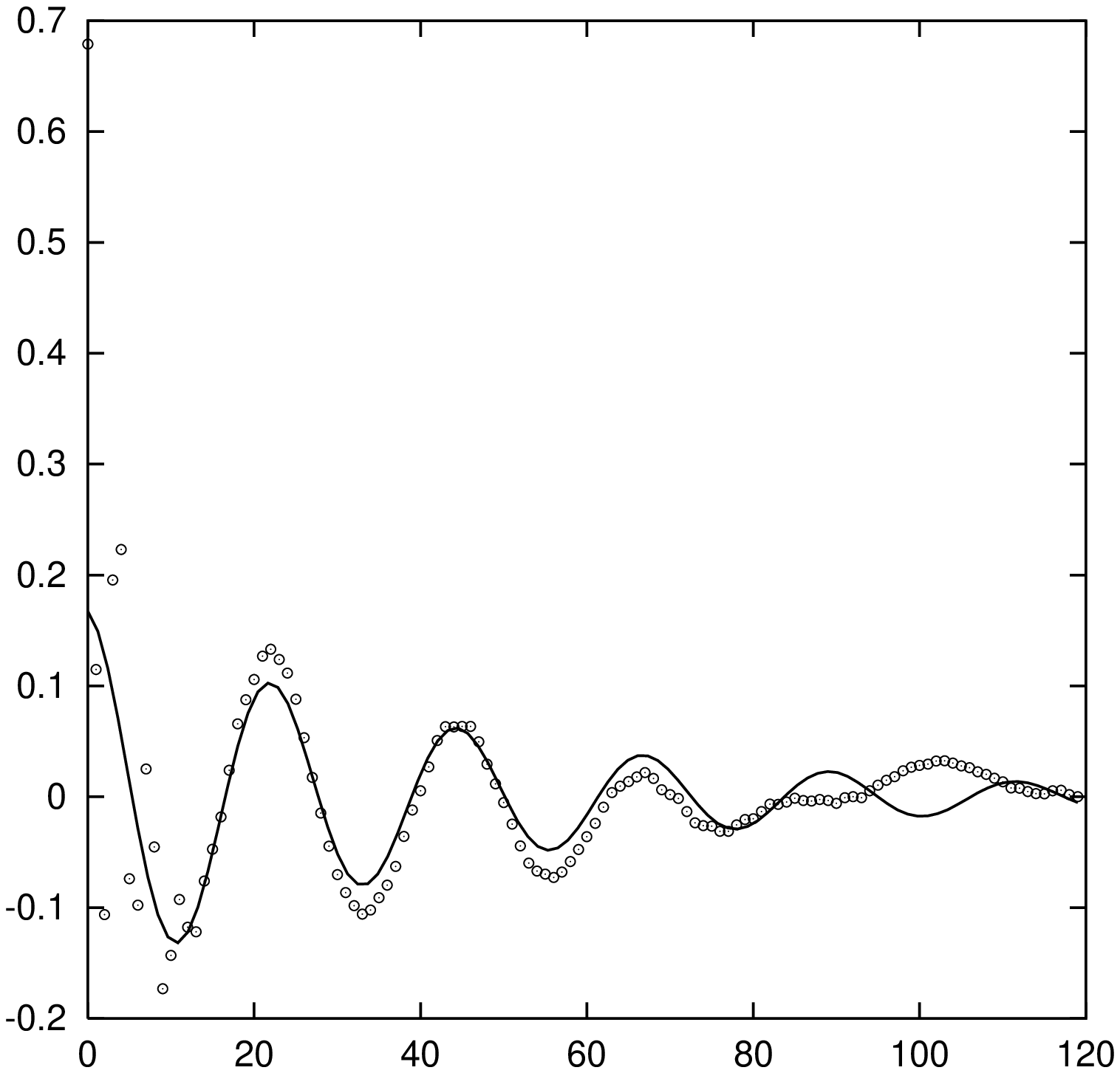}}
\put(118,0){{\Large $m$}}
\put(130,2){{\Large \vector(1,0){15}}}
\put(-1,110){{\Large $\gamma_m$}}
\put(3,120){{\Large \vector(0,1){15}}}
\end{picture}
\caption{Correlation function for $\tau_b = 25\%$ and sequence length
$N=120$. The $g(m)$ plot corresponds to $A=0.17$, $\xi = 45 \pm 6$, $d=22.4 \pm 0.3$
and $\phi=90$.}
\label{fig3}
\end{figure}

For each figure, we plotted both the simulation results and a
theoretical function that approximates these results. This function is
given by

\begin{eqnarray}
g(m) = A \times \exp(\frac{-m}{\xi}) \sin(\frac{2\pi m}{d} +\phi)
\label{eq:correlation}
\end{eqnarray}

where $A, \xi, d$, and $\phi$ are fitting parameters.
Here $A$ is the (uninteresting) normalisation constant, $\xi$ is
the correlation length, $d$ is the period of periodicity and $\phi$
is a phase shift.

Figures \ref{fig2},\ref{fig3},\ref{fig4} and \ref{fig5} plot the
correlation function obtained by using more than 2,500 different
sequences of length 120 chosen randomly from an ensemble of 25,000.
The only difference between these 4 experiments is the hydrophobicity
ratio. We distinguish three different regions in each graph. 1)
initial region: the results are affected by two phenomena - the
discretisation of the simulation space onto a lattice and the effect
of painting which will be considered later. 2) the central region:
results are stable and coincide with equation \ref{eq:correlation}. 3)
Final region: this is characterized by a lack of correlations. Since
the chains are open the ends tend to be more hydrophilic, clearly
shown in figure 5.

\subsection{The Effect of Painting}

The oscillations and their amplitudes are controlled by the parameters
$\xi$ and $A$ respectively, and $d$ controls the period. These three
parameters depend on the hydrophobicity ratio $\tau_b$, and the
sequence length $N$. We have determined the function corresponding to
each parameter, as well as their physical relation to the painting
method applied to folded sequences. 
In what follows we analyze the results of the painting
method and demonstrate its limitations.

In a statistically significant way, it was shown that the amino acid
sequences in proteins differ from what is expected from random
sequences. The results of this study based upon real protein sequences
in the SWISS-PROT data base can be found in \cite{8,9}. Our study confirm
the non-random distribution of hydrophobic-hydrophilic monomers along
the chain, and indicate that part of that correlation may be due to an
implied geometry of the condensed globule.

\setlength{\unitlength}{0.62mm}
\begin{figure} 
\vspace{0.02cm}
\begin{picture}(160,160)
\thicklines
\epsfxsize=90mm
\put(0,5){\epsffile{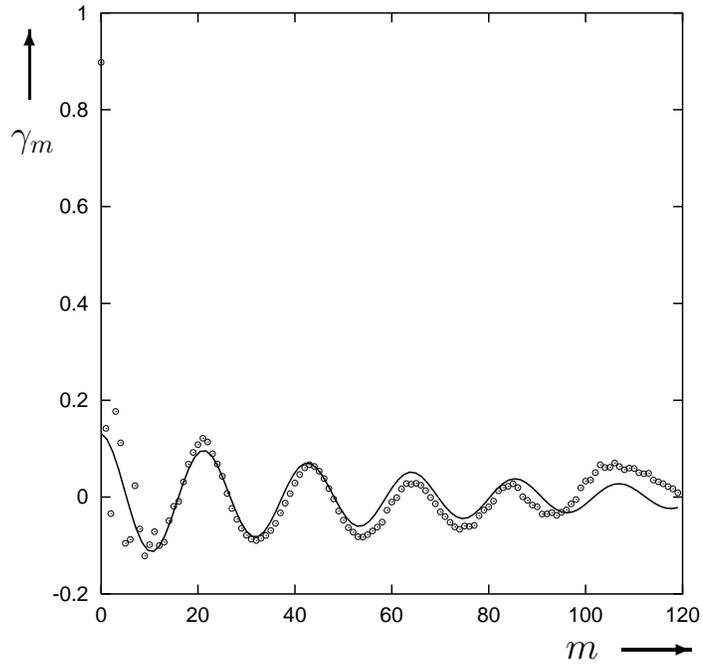}}
\put(118,0){{\Large $m$}}
\put(130,2){{\Large \vector(1,0){15}}}
\put(-1,110){{\Large $\gamma_m$}}
\put(3,120){{\Large \vector(0,1){15}}}
\end{picture}
\caption{Correlation function for $\tau_b = 50\%$ and sequence length
$N=120$. The $g(m)$ plot corresponds to $A=0.13$, $\xi = 69 \pm 11 $, $d=21.4 \pm 0.2 $
and $\phi=90$.}
\label{fig4}
\end{figure}

It is well known that the folding process brings residues
geometrically close together which are also close along the chain.
These residues are classified as domains. The current painting method
does not take domains into account. Quite interesting results for the
monomer structure of folded sequences are obtained applying the method
to short sequences with an appropriate hydrophobicity ratio (Figure
\ref{fig3}). In the case shown in figure \ref{fig3} the domain
structure does not get overwhelmed by the large globule volumes.

When the domain structure is very small with respect to the globule
volume only those monomers at the border of the two volumes (global
and hydrophobic) are considered (see Figure \ref{fig6}). This border
is negligible compared to the rest of the globule. It is vital to take
the domain structure into account to extend the painting method beyond
the globule core.

\subsection{Improvement}

To overcome the limitations of the preceding method we must therefore
take the domain structure into account. We proceed as follows:

\begin{enumerate}
  \item let $\tau$ and $N$ be the hydrophobicity ratio and the
  	sequence length respectively.
  \item choose a point $P_0$ belonging to the globule such that the
  	distance from $P_0$ to the surface of the globule is bigger
  	than $\tau$.
  \item mark the monomers belonging to the volume of radius $\tau$ and
  	centre $P_0$, and derive the characteristics of the folding
  	structure of the monomers (using the correlation function).
  \item repeat steps 2 and 3 with a new point until all possible
  	points have been chosen.
  \item calculate the total correlation function from the preceding
	ones.
\end{enumerate}

This new technique allows the analysis of the monomer folding
structure  for each domain structure in the whole globule. However it
is very time consuming, since all points which satisfy the
hydrophobicity radius $\tau$ must be examined. The method uses the
same exploration procedure as the neural network windowing method.

\setlength{\unitlength}{0.62mm}
\begin{figure} 
\vspace{0.02cm}
\begin{picture}(160,160)
\thicklines
\epsfxsize=90mm
\put(0,5){\epsffile{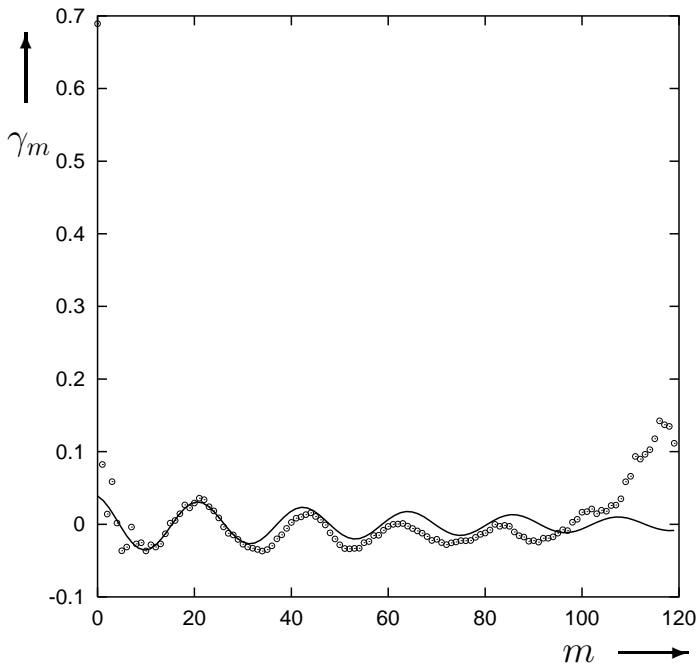}}
\put(118,0){{\Large $m$}}
\put(130,2){{\Large \vector(1,0){15}}}
\put(-1,110){{\Large $\gamma_m$}}
\put(3,120){{\Large \vector(0,1){15}}}
\end{picture}
\caption{Correlation function for $\tau_b = 75\%$ and sequence length
$N=120$. The $g(m)$ plot corresponds to $A=0.04$, $\xi = 77 \pm 25$, $d=21.7 \pm 0.3 $
and $\phi=90$. 
Note that in this figure the quality of fitting is the worst due to the
deep level of painting.
}
\label{fig5}
\end{figure}

\subsection{Neural Network}

Artificial neural networks are usually used to find an approximate
solution to a precisely (or an imprecisely) formulated problem. ANNs
are characterized by the network topology, the connection weight
between pairs of nodes, node properties, and the definition of
updating rules. Usually, an objective function is defined that
represents the complete state of the network, and its set of minima
correspond to different stable states of the network. Learning in an
ANN, whether supervised or unsupervised, is accomplished by adjusting
the weights between connections in response to new inputs or training
patterns.

The advantage of the neural network approach is that it allows us to
generalize our predictions about the compactness of folded proteins
beyond the sequences used to train the network. The results of the
neural network method are fully described in \cite{6}.

\begin{figure} 
\vspace{0.02cm}
\centerline{
\epsfxsize 11cm
\epsffile{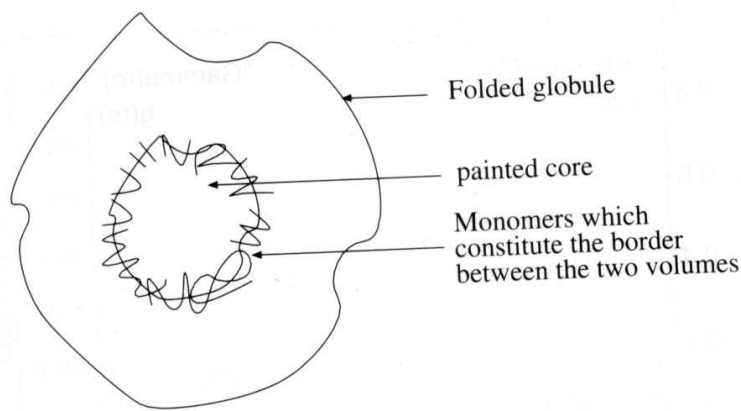}}
\caption{This figure shows the monomers belonging to the border
between the two volumes. This border becomes negligible as the length
of the chain increase.}
\label{fig6}
\end{figure}

\section{Conclusion}

In this paper a method to create geometrical objects with protein-like
structure and thereby generate sequences is suggested. This technique
used two complementary methods - painting method and neural network
windowing method. The first one is suitable to the short sequences. As
opposed to the painting method the ANN windowing method has non-local
effect in the folding process (length of sequences, etc.). However, a
learning process is needed in order to predict the goodness of folded
structure for any sequence length.

We deduce that there are implied chain sequence correlations
indicating a sort of block-like structure to the
hydrophobic-hydrophilic structure. 
We see from the fitting of the experimental function $\gamma_m$
by the theoretical one $g(m)$ that increasing the depth of painting
increases the correlation length $\xi$, while it
has practically no effect on the periodicity $d$ or phase $\phi$.
The periodicity $d$ is believed to be related to the size of the
globule while the phase $\phi$ to the actual procedure involved
and thus both are independent of the painting depth.

The method presented here is very time consuming especially the Monte
Carlo simulation and neural network steps. The performance of these
methods can be improved by using high performance and efficient
parallelization techniques.

\acknowledgments

The authors acknowledge interesting discussions with Dr~A.V.~Gorelov
who was first to experimentally propose the idea of a painted globule
back in 1994 and to Stephen Connolly who performed an undergraduate
project with one of us (E.G.T.) in 1995, during the course of which
the correlation functions of painted globule were first though
yet fairly inaccurately determined from Monte Carlo simulation.


\begin{thebibliography}{99}

\bibitem{1} ~K.~Binder, Ed., 
 \TITLE{Monte Carlo Methods in Statistical Physics, 2nd ed., 
 Springer--Verlag, Berlin (1986)}.

\bibitem{2} ~K.~Binder, Ed.,
 \TITLE{Applications of Monte Carlo Method in Statistical Physics, 2nd ed.,
  Springer--Verlag, Berlin (1987)}.

\bibitem{3} ~M.P.~Allen, D.J.~Tildesley, 
\TITLE{Computer Simulations of Liquids, Clarendon Press, Oxford (1987)}.
 
\bibitem{4} ~A.~Byrne, P.~Kiernan, D.~Green, K.A.~Dawson, 
\IN{J. Chem. Phys.}{102}{1995}{573}.

\bibitem{5} ~Yu.~A.~Kuznetsov, E.~G.~Timoshenko, K.~A.~Dawson, 
\IN{J. Chem. Phys.}{103}{1995}{4807}.

\bibitem{6} ~R.G.~Reilly, Kechadi M-T.,  K.A. Dawson, Yu. A. Kuznetsov 
and E.G. Timoshenko,
\IN{ Il Nuovo Cimento.}{20D (12bis)}{1998}{2565}.

\bibitem{8} ~A. Irback, C. Peterson and F. Potthast,
\IN{Physical Review E}{55}{1997}{860}.


\bibitem{9} ~A. Bairoch and B. Boeckmann,
\IN{Nucleic Acids Res.}{22}{1994}{3578}.

\end{thebibliography}
\end{document}